\documentclass[conference]{IEEEtran}
\IEEEoverridecommandlockouts
\usepackage{times}
\usepackage{cite}
\usepackage{comment}
\usepackage{amsmath,amssymb,amsfonts}
\usepackage{graphicx}
\usepackage[table]{xcolor}
\usepackage{multirow}
\usepackage{booktabs}
\usepackage{algorithm}
\usepackage{algpseudocode}
\usepackage{textcomp}

\usepackage{soul}
\usepackage{pgfplots}

\pgfplotsset{compat=1.18}
\usetikzlibrary{arrows.meta}

\usepackage{xcolor}

\definecolor{tabheader}{HTML}{D5A6BD}
\definecolor{funcblue}{HTML}{9FC5E8}
\definecolor{treegreen}{HTML}{B6D7A8}
\definecolor{rulepurple}{HTML}{B4A7D6}
\definecolor{bayesyellow}{HTML}{FFE599}
\definecolor{lazyorange}{HTML}{F9CB9C}
\definecolor{bracketwin}{HTML}{63BE7B}

\makeatletter
\newcommand{\linebreakand}{%
  \end{@IEEEauthorhalign}\hfill\mbox{}\par
  \mbox{}\hfill\begin{@IEEEauthorhalign}}
\makeatother

\begin{document}

\title{Power, Performance, and Area of HPC-Based Malware Detectors:\\ An FPGA and Open-PDK ASIC Study}

\author{
\IEEEauthorblockN{\small Alireza Abolhasani Zeraatkar$^{1}$, Parnian Shabani Kamran$^{1}$, Parsa Mirfasihi$^{2}$, Saleh Khalaj Monfared$^{3}$, Hussain Al-Asaad$^{1}$}
\IEEEauthorblockA{
$^{1}$Electrical and Computer Engineering Department, University of California, Davis, United States\\
$^{2}$Electrical and Computer Engineering Department, San Francisco State University, San Francisco, United States\\
$^{3}$Electrical and Computer Engineering Department, Worcester Polytechnic Institute, Worcester, United States\\
Email: \{abolhasani, pkamran\}@ucdavis.edu, smirfasihi@sfsu.edu, skmonfared@wpi.edu, hsalasaad@ucdavis.edu}
}

\maketitle

\begin{abstract}
Hardware malware detectors (HMDs) classify programs as benign or malicious based on hardware performance counter (HPC) data. Previous studies have shown that ensemble classifiers outperform other classifiers on detection metrics. However, the area, power, and timing costs of HMDs remain open questions. We address these questions across a comprehensive design space comprising multiple classifier families, variable HPC sets, and ensemble learning methods. We designed a pipeline to reconstruct selected trained models, generate synthesizable register-transfer-level (RTL) implementations, validate each design against a bit-accurate golden model, and evaluate the implementations. The results show that tree- and rule-based detectors achieve the highest detection performance with nearly an order of magnitude less area and several times lower energy than function-based classifiers. Reducing the number of HPC inputs to fit a processor's concurrent-counter limit preserves detection performance while substantially reducing area, power, and delay.
\end{abstract}


\section{Introduction}
Malware remains one of the most persistent threats across computing platforms, from desktop and server systems to embedded and IoT devices \cite{zhou2021,zhou2018}. Software-based detection solutions struggle against polymorphic and metamorphic malware variants because malware can use dynamic code alteration and evasion tactics. Antivirus software remains vulnerable to the evasion techniques it is designed to tackle. A decade of research has therefore turned to hardware-assisted detection, based on the assumption that although an attacker can modify software at any time, controlling the hardware is infeasible \cite{basu2026}.

Modern processors contain built-in HPCs that observe microarchitectural events as a program executes. HPCs measure software performance during execution. They monitor a wide range of events, including instructions, cache hits and misses, branch mispredictions, and TLB misses. Demme et al. showed that malware leaves recognizable fingerprints in HPC data \cite{demme2013}. Subsequent studies have developed classifiers that monitor HPCs continuously by reading a subset of counters and flagging each application as benign or malicious \cite{basu2026,ozsoy2015}. One design constraint for an HMD is the number of HPCs available to count microarchitectural events. The maximum number of counters that can be read concurrently is set at processor design time. For example, some Intel and AMD processors can read only four and six HPCs concurrently, respectively \cite{zhou2018}; therefore, a deployable detector must fit within the processor's counter limit.

Prior studies compare classifiers using reduced HPC configurations. The study in \cite{zeraatkar2024} evaluates 144 configurations and finds that ensemble methods, particularly AdaBoost \cite{freund1997decision} and bagging \cite{breiman1996bagging}, achieve high detection performance. However, these studies leave a key hardware-design question unanswered:

\emph{What are the power, performance, and area costs of each HMD?}

An HMD's area, power, and delay determine its integration overhead to the processor. High-performance ranked HMDs that disregard these parameters will affect the processor performance and efficiency. As a result, a designer needs to prioritize a detector that fits the design space of a processor. Prior works are limited to a few integrated classifier prototypes \cite{ozsoy2015} or FPGA synthesis of a small classifier set  \cite{khasawneh2015} \cite{khasawneh2020ehmd}. No prior study has measured hardware costs across a comprehensive design space for both FPGA and ASIC targets.

This paper evaluates the power, performance, and area of 15 of the 18 classifiers in \cite{zeraatkar2024}. Each classifier is trained with 16, 8, 4, or 2 HPCs; AdaBoost and bagging ensembles are also evaluated with 4 or 2 counters, for a total of 120 configurations. Three research questions (RQs) structure this study.

\textbf{RQ1:} For each classifier family (trees, rules, and function-based models), what are the routed ASIC area, energy per inference, and critical-path delay? What FPGA logic resources and frequency does an RTL implementation require?

\textbf{RQ2:} When the feature set is reduced from 16 to 8, 4, and 2 counters, how much area and energy does each classifier save, how much does its $F_1$ score drop, and which configuration provides the best balance?

\textbf{RQ3:} At the same HPC budget, do AdaBoost and bagging ensembles improve $F_1$ score enough to justify their additional area and energy costs?

This paper makes four contributions to the analysis of trade-offs between detection performance and hardware cost. We developed an automated pipeline to reconstruct selected trained models from the experiments reported in~\cite{zeraatkar2024}, generate synthesizable Verilog RTL, validate each design against a bit-accurate golden model, and characterize completed implementations on FPGA and ASIC targets.

\begin{itemize}

\item We developed a machine-learning-to-silicon flow that covers the detector configurations evaluated in the previous study.

\item We characterized implementation costs using successful FPGA place-and-route results and completed ASIC implementation runs.

\item We analyze $F_1$ scores to explain differences in the hardware characteristics of the classifier families.

\item We recommend a classifier and HPC configuration for the evaluated counter-limited design space.
\end{itemize}
The rest of this paper is organized as follows. Section~\ref{sec:related} reviews HPC-based malware detection, prior implementation-cost studies of HMD classifiers, alternative detection channels, evasion, and the translation of trained models into hardware. Section~\ref{sec:method} describes the experimental setting, design space, model-to-silicon pipeline and validation, and FPGA and ASIC flows. Section~\ref{sec:results} presents detection metrics and implementation costs. Section~\ref{sec:limits} discusses measurement limitations and directions for future work, and Section~\ref{sec:conclusion} concludes the paper.

\section{Related Work}
\label{sec:related}

\subsection{Foundations of HPC-based malware detection.}
Demme et al.~\cite{demme2013} first detected malware online from existing HPCs at roughly 90\% accuracy, based on the observation that malware within a family exhibits similar functional phases regardless of how it is written. Basu et al.~\cite{basu2026} derived the detection probability achievable when counters are sampled at fixed intervals. Ozsoy et al.~\cite{ozsoy2015, ozsoy2016} integrated a detector into the processor pipeline at low overhead, and Khasawneh et al.~\cite{khasawneh2015, khasawneh2020ehmd} showed that ensembles of low-complexity detectors more than halve a single detector's false-positive rate. The study in~\cite{zeraatkar2024} extended that ensemble work to a comprehensive set of configurations ranked by accuracy and area under the receiver operating characteristic curve (AUC), supplying the trained models whose implementation costs this paper measures. Later work has turned to more complex targets such as stealthy malware hidden inside benign programs~\cite{sayadi2021stealth}, zero-day malware recognized by transfer learning on imaged counter traces~\cite{he2024deephmd}, transformer models over tabular counter data~\cite{he2026transformers}, and explainable anomaly detection for medical IoT devices~\cite{golzari2026}. Konstantinou et al.~\cite{konstantinou2022} trace the transition of HMDs from prototypes to products, which makes implementation cost a practical question.

\subsection{Implementation cost of HMD classifiers.}
Ozsoy et al.'s study~\cite{ozsoy2015} is the most relevant to our work; however, it reports only one integrated design point. Khasawneh et al.~\cite{khasawneh2020ehmd} synthesized their ensemble detectors on an FPGA and reported minimal overhead. Makrani et al.~\cite{makrani2022optiedge} estimated the resources and timing of detection classifiers under high-level synthesis. Gao et al.~\cite{gao2021adaptive} and He et al.~\cite{he2022reinforced} select among detectors at run time by weighing detection performance against hardware cost. Beyond the classifier itself, PHMon~\cite{delshadtehrani2020} implements a programmable hardware monitor with an area overhead of 13.5\% and a power overhead of 5\%. However, no prior study evaluates a complete design space. Existing studies consider only a small number of classifiers on a single hardware target and do not verify the functional equivalence of the hardware implementations to the trained models. The design space evaluated in~\cite{zeraatkar2024} spans three orders of magnitude in model size. Therefore, conclusions based on only a few configurations may not generalize across the full design space. When hardware cost and detection performance are considered together, a 4-HPC decision tree provides the best trade-off among the evaluated configurations.

\subsection{Granularity of the HPC budget.}
Zhou et al.~\cite{zhou2021, zhou2018} and Kazdagli et al.~\cite{kazdagli2016} in the mobile setting indicate that reported HMD accuracies depend strongly on experimental design. This motivates reproducing the data split and validating the RTL before assigning hardware costs to accuracy claims. Kadiyala et al.~\cite{kadiyala2020} push observation down to the system-call level, while Kuruvila et al.~\cite{kuruvila2021} propose custom-designed counters. Both approaches address the concurrent-counter limit that shapes RQ2~\cite{zhou2018}, making the HPC budget a design variable.

\subsection{Alternative channels and narrowed targets.}
Agredo et al.~\cite{agredo2025b} detect malware at up to 81\% binary accuracy from TLB counters alone, without cache, branch-predictor, or timing features. The instruction-TLB miss event is among the 16 events we profile, although cache accesses dominate after correlation-based reduction~\cite{zeraatkar2024}. Other studies narrow the target by detecting ransomware with autoencoders~\cite{alam2020}, monitoring before encryption~\cite{anand2023}, analyzing imaged traces~\cite{ganfure2023}, or deploying detectors in IoT environments~\cite{mofidi2025}. Machine-learning detectors that run in software pursue the same threats using other evidence. Prior studies analyze file-level ransomware features~\cite{majd2023ransomware}, classify obfuscated ransomware families~\cite{cassel2023ransomware}, deploy lightweight models for detecting obfuscated malware on resource-constrained IoT devices~\cite{cassel2024iot}, and inspect network traffic at the IoT edge~\cite{radhakrishna2024edge}.
An HMD complements them by observing each program's execution on the processor itself.

\subsection{Evasion, adversarial resilience, and faults.}
Khasawneh et al.~\cite{khasawneh2017rhmd} showed that deployed HMDs can be reverse-engineered and evaded, and proposed stochastic switching among a diverse pool of classifiers. Islam et al.~\cite{islam2022} model switching among multiple classifiers as a Bayesian Stackelberg game. A fully parallel hardware implementation of this defense requires additional resources because it instantiates several classifiers simultaneously. As the detector pool grows, its area and leakage power generally increase. The per-detector measurements reported here provide a basis for estimating this overhead, although shared logic and routing may prevent perfectly linear scaling. He et al.~\cite{he2024obfusgate} show that code obfuscation degrades HMD performance and that representation learning can restore detection. They also use deep reinforcement learning to strengthen detectors against adversarial techniques~\cite{he2024drl}. A deployed detector can also fail. 

Localized voltage-fault injection~\cite{khojasteh2026} offers an adversary a way to induce such faults deliberately. It also finds that hardening the structural units where criticality concentrates can limit the damage from a single fault, at an additional area cost. Two surveys~\cite{chenet2024, sayadi2024} review hardware-based and HPC-based detection as a whole.

To our knowledge, this is the first study to evaluate detection performance and hardware cost across a complete published classifier design space. It also translates the classifiers into validated RTL and characterizes their implementation costs on an FPGA and through a fully routed open-PDK ASIC flow.
Prior hardware evaluations examine either a single design point~\cite{ozsoy2015} or a small set of FPGA-synthesized detectors~\cite{khasawneh2020ehmd}. Evaluating a larger design space also reveals an implementation limit: some configurations cannot be routed with the limited number of metal layers in the SkyWater 130-nm process technology~\cite{skywater2020}. These routing failures provide evidence about the scalability of the evaluated hardware architectures.
\section{Methodology}
\label{sec:method}

\subsection{Experimental Setting and Detector Design Space}
\label{sec:platform}
Our study is based on the trained classifiers of \cite{zeraatkar2024}. The dataset was collected by observing benign and malicious applications running in containers with \texttt{perf} and contains 3,700 samples per class. It was split 70/30 into training and test sets. Correlation clustering and information-gain ranking divided the 16 HPC events into nested sets of 8, 4, and 2 events. We retain this experimental setting and use the trained models as the starting point for hardware evaluation. The methods described in Sections~\ref{sec:pipeline} and~\ref{sec:asicflow} convert the selected models into synthesizable RTL, verify their predictions, and evaluate their implementations on the FPGA and ASIC, respectively.

The original study evaluated 144 configurations of 18 classifiers. Here, we evaluate 120 configurations of 15 classifiers across four families, using the Weka taxonomy. The tree family includes Random Tree, J48, REPTree, Hoeffding Tree, and Decision Stump. The function family includes the multilayer perceptron (MLP), stochastic gradient descent (SGD), sequential minimal optimization (SMO), Simple Logistic, and Logistic. The rule family consists of JRip, Decision Table, PART, and OneR. The Bayesian family is represented by Bayes Net. These family names follow \cite{zeraatkar2024}. The numerical formats required by each family are discussed in Section~\ref{sec:numfmt}.

We omit three classifier types from this analysis. None of the Random Forest variants fits on the FPGA or completes the 130-nm ASIC flow. Too few of the eight logistic model tree (LMT) variants can be fully implemented to support comparisons across HPC budgets and ensemble types. K-nearest neighbors (KNN) requires its training set to be stored on-chip, which is not feasible in the selected standard-cell flow without memory macros.

\subsection{Model-to-Silicon Pipeline}
\label{sec:pipeline}
The pipeline involves five stages. First, we reconstruct the selected models from the available Weka v3.8.6 classifier outputs. Second, each model is transformed into an intermediate representation containing the relevant threshold comparisons, rule ordering, coefficient vectors, neural-network weights, discretization bins, probability tables, ensemble members, and voting weights, according to classifier type. Third, a golden model evaluates this representation using the arithmetic and decision rules intended for the HMD. Fourth, synthesis generates synthesizable Verilog with a parallel combinational datapath between registered inputs and outputs. Each input feature is a raw 32-bit counter value, and the output indicates a benign or malicious classification. The registered input/output wrapper provides register-to-register paths for timing analysis. Finally, each RTL implementation is simulated against its golden model over the complete dataset and submitted to the FPGA and ASIC flows, as described in Sections~\ref{sec:rtlsim} and~\ref{sec:fpgaflow}.

\subsection{Validation of the Experiments}
\label{sec:golden}

This validation compares each RTL implementation with its Weka reference classifier. We reproduced Weka's 70/30 percentage split by shuffling the dataset with a fixed random seed of 1. The resulting test set contains 1,093 clean files and 1,127 malicious files (2,220 instances). We compared each HMD/HPC configuration's predictions with Weka over all 7,400 instances to measure agreement. Five Hoeffding-tree implementations used majority-class leaves instead of Naive Bayes adaptive leaves and achieved 90.2--98.5\% agreement. Twenty-four bagging configurations used discrete class votes instead of averaging member posterior probabilities and achieved 99.6\% median agreement. One SGD-based configuration achieved 99.2\% agreement, with discrepancies on 61 instances associated with the decision of its highest-weight member. Numerical approximations are explained in Section~\ref{sec:numfmt}. Detection metrics in Section~\ref{sec:deploy} are computed from the Weka reference classifiers' confusion matrices on the 2,220-instance test set, including for configurations whose RTL predictions differ from Weka. 

\subsection{Number-Format Policy}
\label{sec:numfmt}

The numerical formats were chosen according to the operations required by each classifier family. The 16 HPC values are non-negative integers, the largest of which requires 28 bits; each datapath therefore uses 32-bit unsigned inputs. For trees, rules, decision tables, Decision Stump, and OneR, fractional thresholds are mapped to the input-data range. This conversion preserves each threshold condition and does not require a multiplier for comparison. It does not eliminate the prediction and ensemble-aggregation differences described in Section~\ref{sec:golden}. Weka normalization is used by SGD, SMO, Logistic, Simple Logistic, and MLP. The golden-model harness evaluates 8-, 12-, 16-, and 20-bit weight formats with wide accumulators. It selects the most compact format whose accuracy differs from the floating-point reference by no more than 0.05 percentage points on the 2,220-sample test set.

\subsection{RTL Simulation}
\label{sec:rtlsim}

All 120 generated designs are simulated using golden-model predictions for the full set of 7,400 cases, and this evaluation covers 888,000 design--input pairs with no mismatches between RTL and the corresponding golden model. Agreement with Weka is tracked separately, and the reported agreement for the quantized function-based implementations is typically above 99.9\%, with approximately 99.95\% agreement for the MLP implementations using the sigmoid lookup table. These numerical results are considered alongside the model-level deviations reported in Section~\ref{sec:golden}.

\subsection{FPGA and ASIC Flows}
\label{sec:fpgaflow}\label{sec:asicflow}\label{sec:metrics}

The FPGA target is an Intel Cyclone V, and each design is processed with Quartus II 15.0 for mapping, fitting, static timing analysis, and power analysis under a 10-ns clock constraint. All counter inputs and the classification output are assigned to virtual pins so that package pin count does not constrain the design. Timing is analyzed from register to register. Designs that exceed the device's limits are retried on the largest Cyclone V GX. In total, 118 of 120 designs are implemented successfully. The 16-HPC MLP and the 4-HPC bagged MLP exceed the available FPGA resources.

The ASIC target uses the SkyWater 130-nm open PDK~\cite{skywater2020} and the sky130hd standard-cell library. Implementations are generated with OpenROAD-flow-scripts~\cite{ajayi2019}, using a container image pinned by digest with OpenROAD 26Q3 and Yosys 0.68. For each ASIC flow, core utilization is set to 35\%, and placement density is set to 0.55. If a design fails, it is retried according to the policy in Algorithm~\ref{alg:retry}. This policy decreases core use to 35\%, 25\%, 18\%, 13\%, 10\%, and 7\%, and thus increases the available core area for a given size cell by a factor of approximately $1.4 \times$ per level, or by $5 \times$ from the first to the last level. Placement density remains 0.20 above the utilization fraction, spreading cells across the enlarged core. The global-routing parameter $\alpha$ is fixed at 0.20 for the first three tiers, then decreases to 0.15, 0.10, and 0.05 to increase routing capacity. If the number of I/O pins is greater than the number supported by the initial die perimeter, the die is sized to accommodate them. Every design uses the same retry policy and retains its inputs across iterations. Designs that exhaust the policy are reported with their failing stage.

For each ASIC design, we record standard-cell area, utilization, worst-case setup and hold slack, and power. For each FPGA design, we record ALMs, registers, DSP blocks, memory bits, maximum frequency, and dynamic power. 

The implementation area is reported in the native units of each target. FPGAs use ALMs, where configurable logic and registers are embedded within a programmable routing architecture, whereas the ASIC reports the number of standard cells used by the mapped design. Also, FPGA arithmetic uses dedicated DSP blocks whose resource cost is not represented by the ALM count. Therefore, it is not possible to convert ALMs into ASIC cell area using a fixed factor. The results for 4-HPC shown in Table \ref{tab:ppa} illustrate this difference. SGD uses 251 ALMs, less than J48's 398, but its ASIC cell area is 88{,}004 $\mu m^{2}$ as compared to J48's 12{,}731 $\mu m\textsuperscript{2}$. Therefore, for each target the FPGA resources (DSP, memory, and logic) are presented alongside the ASIC area so as to compare the detection--cost trade-offs without assuming identical area rankings.

\begin{algorithm}[t]
\caption{Retry policy for a design that fails physical implementation.}
\label{alg:retry}
\begin{algorithmic}[1]
\Require Verified RTL for design $D$
\Ensure routed implementation of $D$, or $D$ reported with its failing stage
\State $L \gets \big[(35,\,0.55,\,0.20),\ (25,\,0.45,\,0.20),$
\Statex \hspace{1.1em} $(18,\,0.38,\,0.20),\ (13,\,0.33,\,0.15),$
\Statex \hspace{1.1em} $(10,\,0.30,\,0.10),\ (7,\,0.27,\,0.05)\big]$
\Statex \hspace{0.6em}\emph{tiers $k=0\ldots5$ as (utilization $u_k$\,\%, density $\rho_k$, adjustment $\alpha_k$)}
\Statex \hspace{0.6em}$u_k = 35 \cdot 0.725^{k}$ rounded,\ \ $\rho_k = u_k/100 + 0.20$,
\Statex \hspace{0.6em}$\alpha_k = 0.20$ for $k \leq 2$, then lowered by $0.05$ per tier
\If{$D$ has more I/O pins than a tier-0 die perimeter can hold}
  \State append a die sized to its pin perimeter to $L$
\EndIf
\ForAll{$(u,\rho,\alpha) \in L$}
  \State configure the flow with utilization $u$, density $\rho$, adjustment $\alpha$
  \State $r \gets \textsc{RunFlow}(D)$
  \If{$r$ reaches a routed netlist with extracted parasitics}
    \State \Return $r$
  \EndIf
\EndFor
\State \Return $D$ reported with its failing stage
\end{algorithmic}
\end{algorithm}

\section{Implementation Results}
\label{sec:results}

\begin{table*}[!t]
\caption{Power, performance, and area of every 4-HPC single-classifier design point.}
\label{tab:ppa}
\centering\footnotesize\setlength{\tabcolsep}{6pt}
\begin{tabular}{@{}lrrrrrr@{}}
\toprule
 & \multicolumn{3}{c}{ASIC (SkyWater 130\,nm)} & \multicolumn{3}{c}{FPGA (Cyclone V)} \\
\cmidrule(lr){2-4}\cmidrule(l){5-7}
Classifier & Area (\textmu m\textsuperscript{2}) & $t_{crit}$ (ns) & E/inf.\ (pJ) & ALMs & $f_{\max}$ (MHz) & $P_{dyn}$ (mW) \\
\midrule
Rand.Tree   & 27{,}075 &  5.09 & 56.8 &   930 & 100.4 &  2.6 \\
J48         & 12{,}731 &  4.20 & 27.3 &   398 & 105.4 &  1.8 \\
REPTree     & 11{,}124 &  3.67 & 24.6 &   380 & 103.6 &  1.9 \\
Hoeff.Tree  &  1{,}164 &  2.10 &  2.6 &    11 & 187.6 &  0.8 \\
Dec.Stump   &  1{,}225 &  2.09 &  2.9 &    12 & 155.3 &  1.8 \\
\addlinespace[2pt]
JRip        &  7{,}645 &  2.70 & 18.6 &   223 & 128.8 &  1.0 \\
Dec.Table   &  5{,}107 &  3.61 & 10.9 &   267 &  94.7 &  1.4 \\
PART        &  9{,}495 &  3.02 & 21.5 &   317 & 112.1 &  1.0 \\
OneR        &  6{,}248 &  4.15 & 12.8 &   255 & 106.3 &  1.1 \\
\addlinespace[2pt]
MLP         & 193{,}999 & 33.79 & 234.1 & 4{,}603 & 15.4 & 219.7 \\
SGD         & 88{,}004 & 21.31 & 79.4 &   251 &  32.3 & 34.1 \\
Simple Log. & 68{,}820 & 11.00 & 74.1 &   179 &  70.6 & 15.8 \\
SMO         & 93{,}522 & 19.82 & 83.1 &   221 &  34.7 & 33.0 \\
Logistic    & 60{,}935 & 12.61 & 67.0 &   213 &  50.4 & 17.0 \\
\addlinespace[2pt]
BayesNet    & 19{,}997 & 10.32 & 36.8 &   857 &  51.7 &  1.6 \\
\addlinespace[3pt]
\midrule
\multicolumn{7}{@{}l}{\emph{family medians over the rows above}} \\
\addlinespace[1pt]
\emph{Trees}    & 11{,}124 &  3.67 & 24.6 &   380 & 105.4 &  1.8 \\
\emph{Rules}    &  6{,}946 &  3.31 & 15.7 &   261 & 109.2 &  1.1 \\
\emph{Function} & 88{,}004 & 19.82 & 79.4 &   221 &  34.7 & 33.0 \\
\emph{Bayes}    & 19{,}997 & 10.32 & 36.8 &   857 &  51.7 &  1.6 \\
\bottomrule
\end{tabular}
\end{table*}

This section relates the performance of each HMD type and configuration to its hardware cost. Detection metrics come from the Weka reference tests, while resource and timing metrics come from the FPGA and ASIC characterization flows. Power and energy values are estimates derived from those flows, as discussed in Section~\ref{sec:limits}. These metrics describe the reference classifiers and do not imply identical hardware predictions for configurations with documented implementation differences. Results are reported only for configurations that complete all characterization stages.

\subsection{Simulation Setup}
\label{sec:setup}
We used the OpenROAD-flow-scripts \cite{ajayi2019} container \texttt{openroad/orfs} and Yosys 0.68 for logic synthesis and the \texttt{sky130hd} standard-cell library at the (1.8\,V, 25\,$^\circ$C) process corner. Switching activity for power analysis was extracted from gate-level simulations of each routed netlist using the SkyWater \texttt{sky130\_fd\_sc\_hd} cell models~\cite{skywater2020}. The FPGA target is characterized using Quartus~II~15.0. The ASIC flow ran on Amazon EC2 instances with 32 cores of an Intel Xeon Platinum with 64\,GiB of memory, and the FPGA flow ran on machines with 16\,GB of memory.

\subsection{Deployment Metrics}
\label{sec:deploy}
\begin{figure}[t]
\centering
\includegraphics[width=\columnwidth]{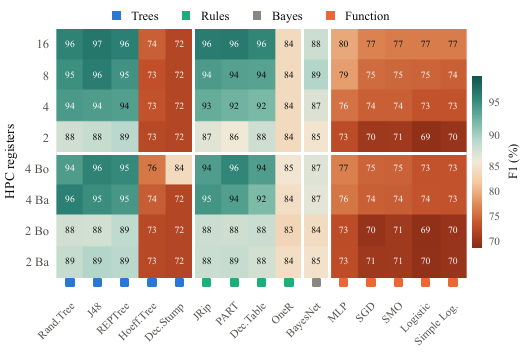}
\caption{$F_1$ scores for all 120 classifier configurations, across HPC budgets and for single classifiers and ensembles.}
\label{fig:f1heatmap}
\end{figure}

\begin{figure*}[t]
\centering
\includegraphics[width=\textwidth]{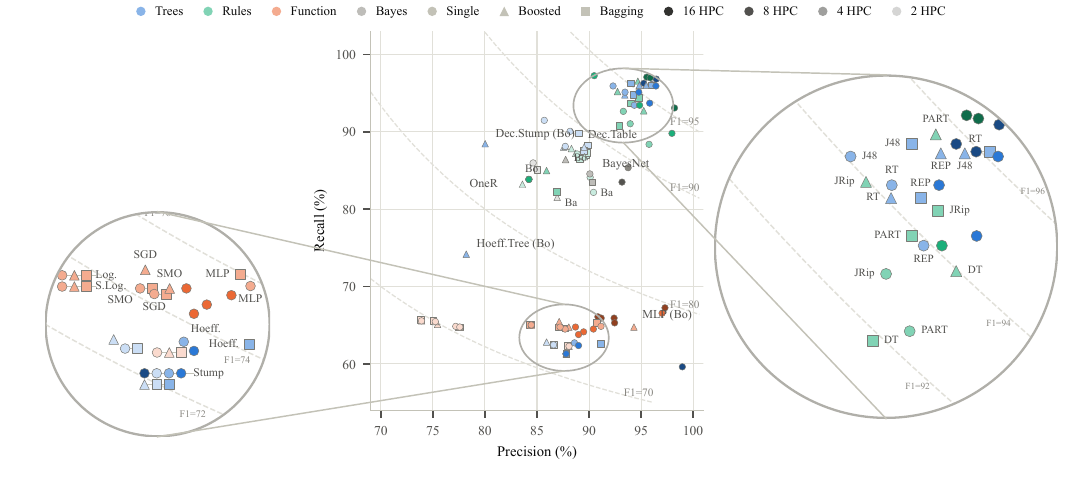}
\caption{Precision against recall for all 120 configurations, with iso-$F_1$ contours.}
\label{fig:pr}
\end{figure*}

Precision, recall, and $F_1$ score provide further information about each HMD configuration. We recalculate these metrics from the confusion matrices and summarize the results in Figs.~\ref{fig:f1heatmap} and~\ref{fig:pr}. In some function-based configurations, recall falls to 62\%, meaning that about 38\% of malicious samples are incorrectly classified as benign. Function-based classifiers perform poorly compared with tree- and rule-based classifiers. With four HPCs, the single Random Tree, J48, REPTree, JRip, PART, and Decision Table classifiers achieve $F_1$ scores of about 92\% to 94\%, compared with 73\% to 76\% for the function-based classifiers. 

The effect of ensembling depends on the classifier and the metric. With four HPCs, selecting the higher-$F_1$ ensemble of AdaBoost and bagging for each classifier yields median improvements of 1.15 percentage points in recall and 0.48 percentage points in precision relative to the corresponding single classifier. The median reduction in false-positive rate is 0.36 percentage points. For J48, AdaBoost increases the $F_1$ score by approximately 1.7 percentage points, mainly through a 3.2-percentage-point increase in precision and a reduction in false-positive rate from 8.2\% to 4.7\%. Recall remains approximately 96\%.

\subsection{Implementation Cost: Power, Performance, and Area}
\label{sec:hw}

\begin{figure*}[t]
\centering
\includegraphics[width=\textwidth]{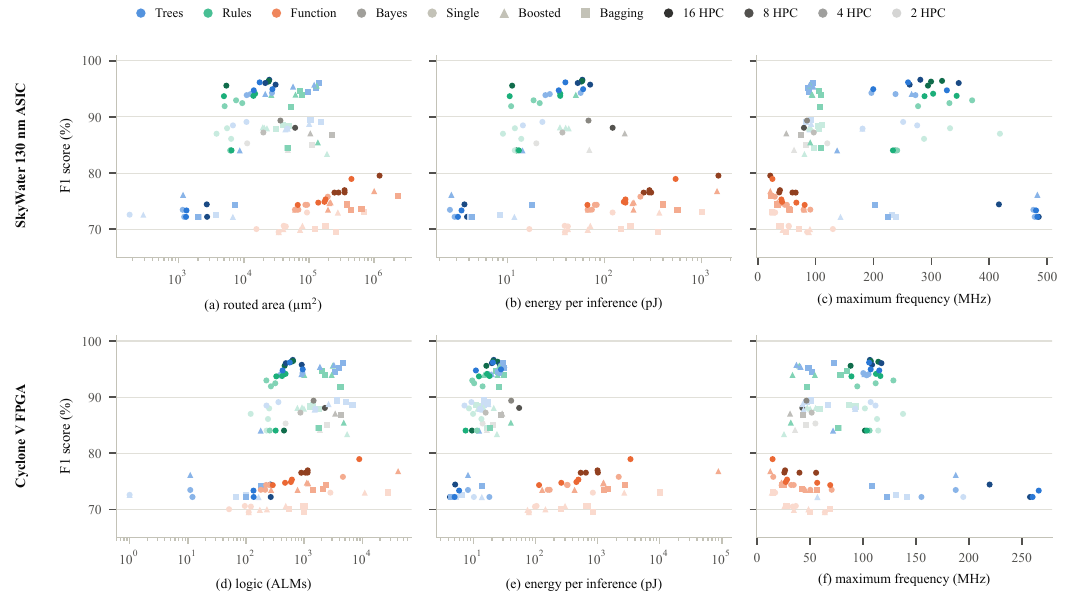}
\caption{Detection performance against implementation metrics: (a, d) area, (b, e) energy per inference, and (c, f) maximum frequency.}
\label{fig:cost}
\end{figure*}

\begin{figure*}[t]
\centering
\includegraphics[width=\textwidth]{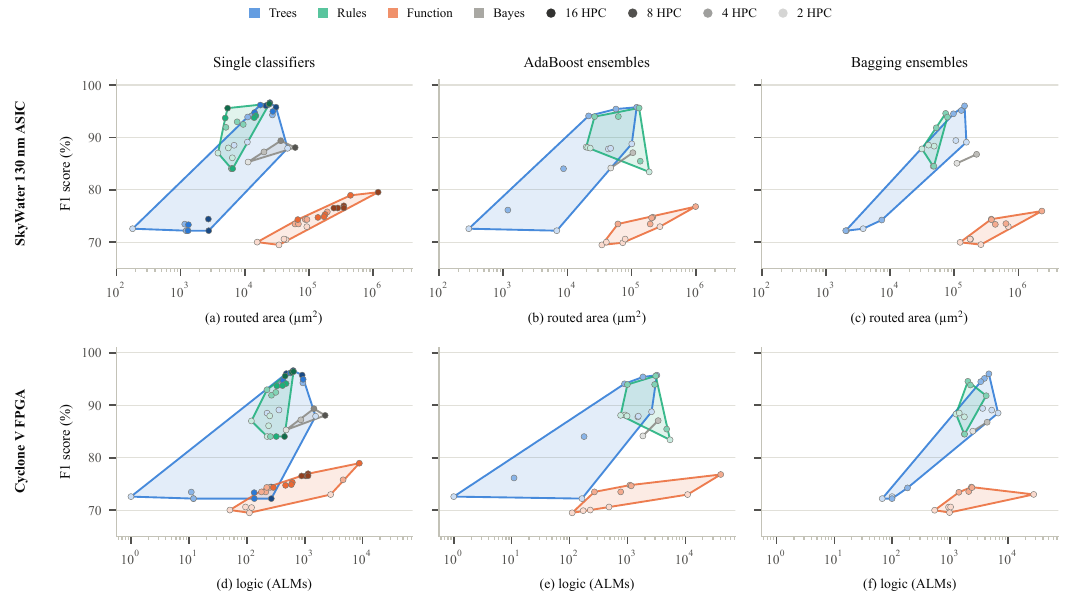}
\caption{Detection performance against implementation cost by ensemble type, with each family's region specified. Routed SkyWater 130-nm ASIC area is shown above, and Cyclone~V ALMs are shown below.}
\label{fig:regions}
\end{figure*}

\begin{figure*}[t]
\centering
\includegraphics[width=\textwidth]{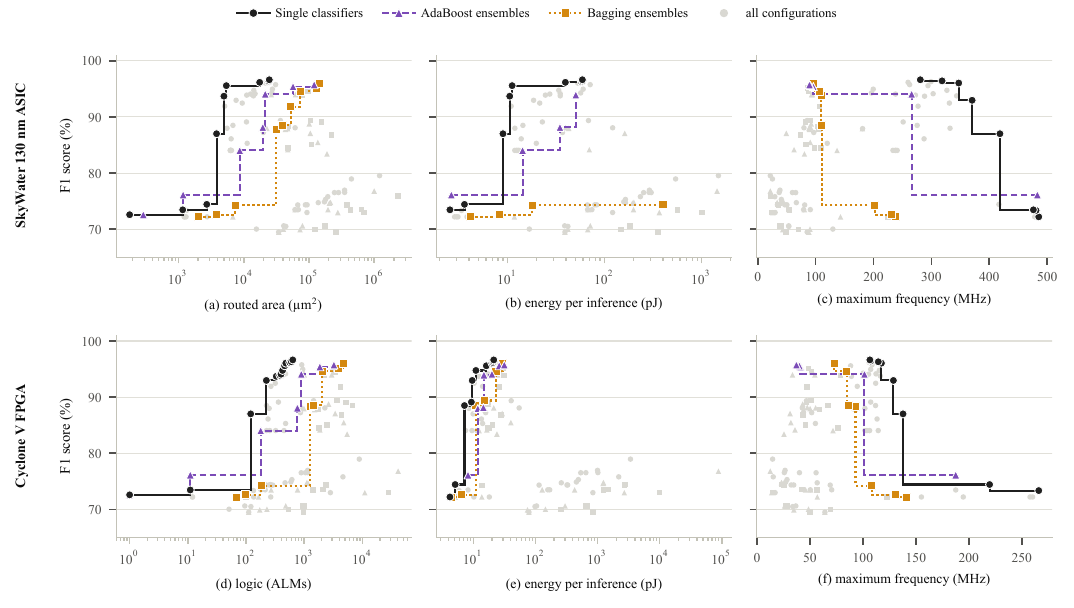}
\caption{Pareto frontiers for single classifiers, AdaBoost, and bagging in each panel of Fig.~\ref{fig:cost}.}
\label{fig:pareto-overlay}
\end{figure*}

The family regions in the figures show why the two hardware targets should be compared separately. On the ASIC, single-function classifiers have a lower $F_1$ but a larger area than strong tree and rule classifiers, but on the FPGA their ALM counts overlap with some tree and rule implementations, because their arithmetic can use dedicated DSP blocks. For example, the 4-HPC SGD implementation uses fewer ALMs than J48 (251 versus 398) but requires substantially more ASIC area (88,004 versus 12,731 $\mu\mathrm{m}^2$). The two targets therefore support similar conclusions about detection-cost trade-offs without implying identical area rankings across classifier families.

Among the four-HPC designs, we recommend J48. Its ASIC implementation has an area of 12,731~$\mu\mathrm{m}^2$, requires an estimated 27.3~pJ per inference, and uses 398 ALMs on the FPGA. More area-efficient alternatives are Decision Table (5,107~$\mu\mathrm{m}^2$), OneR (6,248~$\mu\mathrm{m}^2$), JRip (7,645~$\mu\mathrm{m}^2$), PART (9,495~$\mu\mathrm{m}^2$), and REPTree (11,124~$\mu\mathrm{m}^2$). These five classifiers achieve recall values from 83.8\% to 93.4\%, compared with 95.9\% for J48. Their false-negative rates are approximately 1.6 to 4.0 times J48's rate, in exchange for area savings of up to 7,624~$\mu\mathrm{m}^2$.

Among single classifiers using at most four HPCs, only Random Tree has an $F_1$ score 0.2 percentage points higher than J48, but it requires 2.1 times J48's ASIC area and energy and has lower recall. J48 therefore has the highest reference recall among single classifiers within this HPC budget, with an $F_1$ score close to the best in that subset. Higher $F_1$ scores are possible with a larger resource budget. All single-classifier configurations within one percentage point of the highest $F_1$ use at least eight HPCs; the 8-HPC J48 has the lowest area in that group at 17,549~$\mu\mathrm{m}^2$. The 4-HPC ensembles within the same $F_1$ interval require approximately 9.4 to 11.3 times the ASIC area of the single 4-HPC J48 in this comparison.

The Pareto analysis evaluates each implementation independently. A configuration is nondominated within a comparison set if no other design is at least as good on every metric in that set and strictly better on at least one. Lower area and energy are preferred; higher frequency is preferred in frequency comparisons. These frontiers describe the available plotted results and do not imply that missing ASIC energy results would leave the frontier unchanged.

Tree- and rule-based classifiers dominate most of the reported comparisons. SGD is an exception on the ASIC energy frontier: the lower-energy tree and rule designs with ASIC energy estimates are the Hoeffding Tree and Decision Stump. SGD achieves an $F_1$ score about 0.2 percentage points above the 4-HPC bagged Hoeffding Tree, but uses about 22 times its estimated energy.

Single classifiers generally lie on or near the strongest cost-performance frontiers, although some ensembles improve on their base classifiers. For example, four-HPC AdaBoost improves J48's $F_1$ score by about 1.7 percentage points. The 16-HPC J48 achieves the highest $F_1$ score in the evaluated design space and requires about one-fifth as many FPGA resources as the best bagging and AdaBoost configurations. The boosted four-HPC Hoeffding Tree is a notable exception among ensemble frontier points: its $F_1$ score is close to 76\%, well below those of the stronger tree and rule classifiers.

The 4-HPC J48 lies on the FPGA ALM Pareto frontier among single classifiers. The 16-HPC Decision Table dominates it in the unrestricted ASIC area and energy comparisons, but requires four times as many input counters. Within the four-HPC budget, J48 balances ASIC area and energy with FPGA ALM usage and maximum frequency. This balance supports recommending J48 when all of these costs matter, rather than selecting a design solely for minimum area or maximum $F_1$.

\section{Limitations and Future Work}
\label{sec:limits}

This study evaluates datapaths with fully parallel, register-controlled inputs and outputs. Because the ASIC flow uses a fixed design without per-classifier optimization, the reported costs are representative of the selected architecture, tool settings, and technology level. 

The ASIC implementations were characterized by their maximum clock frequency. ASIC power and energy were estimated from unit-delay gate-level simulations of shuffled dataset samples at the typical process corner. Uniform cell delays do not capture routed delays or interconnect effects, and shuffled samples do not preserve temporal correlations between successive counter readings during program execution. Therefore, the ASIC power and energy estimates may not represent the consumption of fabricated hardware. FPGA power and energy are also estimates; FPGA energy per inference is computed from dynamic power at the nominal 10-ns clock period. 

Future work could study clock gating between counter samples to reduce unnecessary register and clock activity and its effect on energy per inference. It could also evaluate retiming-driven synthesis and physical implementation to reduce critical-path delay. We will evaluate pipelining to increase inference throughput while accounting for the added register area, power, and latency. These studies could further refine the trade-offs among these design parameters.

\section{Conclusions}
\label{sec:conclusion}
This paper compiles 120 trained HPC-based malware-detector configurations into verified RTL and evaluates their hardware characteristics using an Intel Cyclone V FPGA and SkyWater 130-nm ASIC flow. We report three main findings.

First, certain tree- and rule-based classifiers offer good detection-to-cost ratios on both targets. With four HPCs, the function-based family has 7.9 times the median ASIC area of the tree-based family and 12.7 times that of the rule-based family. The median energy ratios relative to the tree- and rule-based families are 3.2 and 5.1, respectively. These costs are associated with a reduction in the reference detection rate; recall rates on some function-oriented classifier configurations can be as low as 62\%. 

Second, the HPC budget produces a measurable area--detection trade-off. Reducing J48's budget from 16 to 4 HPCs lowers area by 48.3\%, while its $F_1$ score decreases from 96.63\% to 94.08\%. Reducing the budget further to 2 HPCs saves another 46.2\% of area but lowers $F_1$ to 88.50\%. These results support four counters as a favorable operating point for J48 within the evaluated configurations.

Third, ensemble improvements are costly in terms of hardware resources. When using four HPCs, AdaBoost improves J48's $F_1$ by approximately 1.7 percentage points, which is mainly due to increased precision and reduced false positive rates. The 4-HPC ensembles within one percentage point of the best $F_1$ in the design space require about 9.4 to 11.3 times the ASIC area of the single 4-HPC J48. 

The resulting recommendation is a single 4-HPC J48 detector for the evaluated counter-constrained setting. Its implementation occupies 12,731~$\mu\mathrm{m}^2$ and requires an estimated 27.3~pJ per inference on the ASIC, while using 398 ALMs on the FPGA. Its reference $F_1$ of 94.08\% is approximately 98\% of the highest ensemble $F_1$, and its 95.9\% recall is the highest among the evaluated single classifiers using at most four HPCs.

\section*{Acknowledgment}
The authors thank the High Performance Computing Core Facility at the University of California, Davis, for providing computational resources that supported this work. OpenAI Codex (GPT-6) assisted with proofreading, language editing, and consistency checks across the Abstract, Sections I–VI, and Acknowledgment. It also assisted with limited citation and reference-formatting corrections. It was not used to generate the study’s research design, equations, data, results, or scientific conclusions. The authors are responsible for reviewing and approving the final manuscript.


\begin{thebibliography}{00}
\bibitem{zhou2021} B. Zhou, A. Gupta, R. Jahanshahi, M. Egele, and A. Joshi, ``A cautionary tale about detecting malware using hardware performance counters and machine learning,'' IEEE Design Test, vol. 38, no. 3, pp. 39-50, 2021.
\bibitem{zhou2018} B. Zhou, A. Gupta, R. Jahanshahi, M. Egele, and A. Joshi, ``Hardware Performance Counters Can Detect Malware: Myth or Fact?,'' in Proc. of the 2018 Asia Conference on Computer and Communications Security (ASIACCS '18), Incheon, Republic of Korea, pp. 457-468, 2018.
\bibitem{basu2026} K. Basu, P. Krishnamurthy, F. Khorrami, and R. Karri, ``A Theoretical Study of Hardware Performance Counters-Based Malware Detection,'' IEEE Transactions on Information Forensics and Security, vol. 15, pp. 512-525, 2020.
\bibitem{demme2013} J. Demme et al., ``On the feasibility of online malware detection with performance counters,'' ACM SIGARCH Computer Architecture News, vol. 41, no. 3, pp. 559-570, 2013.
\bibitem{ozsoy2015} M. Ozsoy, C. Donovick, I. Gorelik, N. Abu-Ghazaleh, and D. Ponomarev, ``Malware-aware processors: A framework for efficient online malware detection,'' in Proc. of the 2015 IEEE 21st International Symposium on High Performance Computer Architecture (HPCA), pp. 651-661, 2015.
\bibitem{zeraatkar2024} A. Abolhasani Zeraatkar, P. Shabani Kamran, I. Kaur, N. Ramu, T. Sheaves, and H. Al-Asaad, ``On the performance of malware detection classifiers using hardware performance counters,'' in Proc. of the 2024 International Conference on Smart Applications, Communications and Networking (SmartNets), pp. 1-6, 2024.
\bibitem{freund1997decision} Y. Freund and R. E. Schapire, ``A decision-theoretic generalization of on-line learning and an application to boosting,'' in Journal of Computer and System Sciences, vol. 55, pp. 119-139, 1997.
\bibitem{breiman1996bagging} L. Breiman,  ``Bagging predictors,'' Machine Learning, vol. 24, no. 2, pp. 123--140, Aug. 1996.
\bibitem{khasawneh2015} K. N. Khasawneh, M. Ozsoy, C. Donovick, N. Abu-Ghazaleh, and D. Ponomarev, ``Ensemble learning for low-level hardware-supported malware detection,'' in Proc. of the 18th International Symposium on Research in Attacks, Intrusions, and Defenses (RAID), Lecture Notes in Computer Science, vol. 9404, pp. 3-25, 2015.
\bibitem{khasawneh2020ehmd} K. N. Khasawneh, M. Ozsoy, C. Donovick, N. Abu-Ghazaleh, and D. Ponomarev, ``EnsembleHMD: Accurate hardware malware detectors with specialized ensemble classifiers,'' IEEE Transactions on Dependable and Secure Computing, vol. 17, no. 3, pp. 620-633, 2020.
\bibitem{ozsoy2016} M. Ozsoy, K. N. Khasawneh, C. Donovick, I. Gorelik, N. Abu-Ghazaleh, and D. Ponomarev, ``Hardware-based malware detection using low-level architectural features,'' IEEE Transactions on Computers, vol. 65, no. 11, pp. 3332-3344, 2016.
\bibitem{sayadi2021stealth} H. Sayadi et al., ``Towards accurate run-time hardware-assisted stealthy malware detection: A lightweight, yet effective time series CNN-based approach,'' Cryptography, vol. 5, no. 4, art. no. 28, 2021.
\bibitem{he2024deephmd} Z. He, H. Homayoun, and H. Sayadi, ``Guarding against the unknown: Deep transfer learning for hardware image-based malware detection,'' Journal of Hardware and Systems Security, vol. 8, no. 2, pp. 61-78, 2024.
\bibitem{he2026transformers} Z. He and H. Sayadi, ``Transformers for tabular anomaly detection in hardware-assisted security: A systematic and empirical study,'' in Proc. of the Great Lakes Symposium on VLSI (GLSVLSI), pp. 214-219, 2026.
\bibitem{golzari2026} T. Golzari, Z. He, and H. Sayadi, ``Bridging hardware and semantics: An LLM-enhanced ontology framework for explainable anomaly detection in IoMT systems,'' in Proc. of the 2026 International Conference on Semantic Computing (ICSC), pp. 1-8, 2026.
\bibitem{konstantinou2022} C. Konstantinou, X. Wang, P. Krishnamurthy, F. Khorrami, M. Maniatakos, and R. Karri, ``HPC-based malware detectors actually work: Transition to practice after a decade of research,'' IEEE Design Test, vol. 39, no. 4, pp. 23-32, 2022.
\bibitem{makrani2022optiedge} H. M. Makrani, Z. He, S. Rafatirad, and H. Sayadi, ``Accelerated machine learning for on-device hardware-assisted cybersecurity in edge platforms,'' in Proc. of the 2022 23rd International Symposium on Quality Electronic Design (ISQED), pp. 77-83, 2022.
\bibitem{gao2021adaptive} Y. Gao et al., ``Adaptive-HMD: Accurate and cost-efficient machine learning-driven malware detection using microarchitectural events,'' in Proc. of the 2021 IEEE 27th International Symposium on On-Line Testing and Robust System Design (IOLTS), pp. 1-7, 2021.
\bibitem{he2022reinforced} Z. He, H. M. Makrani, S. Rafatirad, H. Homayoun, and H. Sayadi, ``Breakthrough to adaptive and cost-aware hardware-assisted zero-day malware detection: A reinforcement learning-based approach,'' in Proc. of the 2022 IEEE 40th International Conference on Computer Design (ICCD), pp. 231-238, 2022.
\bibitem{delshadtehrani2020} L. Delshadtehrani, S. Canakci, B. Zhou, S. Eldridge, A. Joshi, and M. Egele, ``PHMon: A programmable hardware monitor and its security use cases,'' in Proc. of the 29th USENIX Security Symposium (USENIX Security '20), pp. 807-824, 2020.
\bibitem{kazdagli2016} M. Kazdagli, V. J. Reddi, and M. Tiwari, ``Quantifying and improving the efficiency of hardware-based mobile malware detectors,'' in Proc. of the 49th Annual IEEE/ACM International Symposium on Microarchitecture (MICRO), pp. 1-13, 2016.
\bibitem{kadiyala2020} S. P. Kadiyala, P. Jadhav, S.-K. Lam, and T. Srikanthan, ``Hardware performance counter-based fine-grained malware detection,'' ACM Transactions on Embedded Computing Systems (TECS), vol. 19, no. 5, pp. 1-17, 2020.
\bibitem{kuruvila2021} A. P. Kuruvila, A. Mahapatra, R. Karri, and K. Basu, ``Hardware Performance Counters: Ready-Made vs Tailor-Made,'' ACM Transactions on Embedded Computing Systems (TECS), pp. 1-26, 2021.
\bibitem{agredo2025b} C. Agredo, D. F. Koranek, C. M. Schubert Kabban, J. A. Gutierrez del Arroyo, and S. R. Graham, ``Microarchitectural malware detection via translation lookaside buffer (TLB) events,'' Journal of Cybersecurity and Privacy, vol. 5, no. 3, art. no. 75, 2025.
\bibitem{alam2020} M. Alam, S. Sinha, S. Bhattacharya, S. Dutta, D. Mukhopadhyay, and A. Chattopadhyay, ``Rapper: Ransomware prevention via performance counters,'' arXiv preprint arXiv:2004.01712.
\bibitem{anand2023} P. Mohan Anand, P. V. Sai Charan, and Sandeep K. Shukla, ``HiPeR - Early Detection of a Ransomware Attack using Hardware Performance Counters,'' Digital Threats, vol. 4, no. 3, pp. 43, Sep. 2023.
\bibitem{ganfure2023} G. O. Ganfure, C.-F. Wu, Y.-H. Chang, and W.-K. Shih, ``DeepWare: Imaging Performance Counters With Deep Learning to Detect Ransomware,'' IEEE Trans. Comput., vol. 72, no. 3, pp. 600-613, Mar. 2023
\bibitem{mofidi2025} F. Mofidi, S. G. Hounsinou, and G. Bloom, ``RansomSentry: Hardware-level monitoring for ransomware protection in IoT environments,'' in Proc. of the 2025 IEEE International Conference on Pervasive Computing and Communications Workshops and other Affiliated Events (PerCom Workshops), 2025.
\bibitem{majd2023ransomware} N. Ebrahimi Majd and T. Mazumdar, ``Ransomware classification using machine learning,'' in Proc. of the 2023 32nd International Conference on Computer Communications and Networks (ICCCN), pp. 1-7, 2023.
\bibitem{cassel2023ransomware} W. Cassel and N. Ebrahimi Majd, ``Obfuscated ransomware family classification using machine learning,'' in Proc. of the 2023 International Conference on Computational Science and Computational Intelligence (CSCI), pp. 788-792, 2023.
\bibitem{cassel2024iot} W. Cassel and N. Ebrahimi Majd, ``A lightweight obfuscated malware multi-class classifier for IoT using machine learning,'' in Proc. of the 2024 International Conference on Computing, Networking and Communications (ICNC), pp. 239-243, 2024.
\bibitem{radhakrishna2024edge} T. Radhakrishna and N. Ebrahimi Majd, ``Edge computing ransomware detection in IoT using machine learning,'' in Proc. of the 2024 International Conference on Computing, Networking and Communications (ICNC), pp. 244-248, 2024.
\bibitem{khasawneh2017rhmd} K. N. Khasawneh, N. Abu-Ghazaleh, D. Ponomarev, and L. Yu, ``RHMD: Evasion-resilient hardware malware detectors,'' in Proc. of the 50th Annual IEEE/ACM International Symposium on Microarchitecture (MICRO), pp. 315-327, 2017.
\bibitem{islam2022} M. S. Islam, K. N. Khasawneh, N. Abu-Ghazaleh, D. Ponomarev, and L. Yu, ``Efficient Hardware Malware Detectors That are Resilient to Adversarial Evasion,'' IEEE Transactions on Computers, vol. 71, no. 11, pp. 2872-2887, 2022.
\bibitem{he2024obfusgate} Z. He, C. W. Fernandes, and H. Sayadi, ``ObfusGate: Representation learning-based gatekeeper for hardware-level obfuscated malware detection,'' in Proc. of the 2024 Design, Automation \& Test in Europe Conference \& Exhibition (DATE), pp. 1-2, 2024.
\bibitem{he2024drl} Z. He, H. Homayoun, and H. Sayadi, ``Beyond conventional defenses: Proactive and adversarial-resilient hardware malware detection using deep reinforcement learning,'' in Proc. of the 61st ACM/IEEE Design Automation Conference (DAC), pp. 1-6, 2024.
\bibitem{khojasteh2026} F. Khojasteh Dana, S. Khalaj Monfared, H. Okhravi, and S. Tajik, ``GlitchSnipe: Toward localized voltage fault attacks,'' Cryptology ePrint Archive, Paper 2026/752, 2026.
\bibitem{chenet2024} C. P. Chenet, A. Savino, and S. Di Carlo, ``A Survey on Hardware-Based Malware Detection Approaches,'' IEEE Access, vol. 12, pp. 54115-54128, 2024.
\bibitem{sayadi2024} H. Sayadi, Z. He, H. M. Makrani, and H. Homayoun, ``Intelligent malware detection based on hardware performance counters: A comprehensive survey,'' in Proc. of the 2024 25th International Symposium on Quality Electronic Design (ISQED), pp. 1-10, 2024.
\bibitem{skywater2020} SkyWater Technology and Google, ``SkyWater open source PDK,'' 2020. [Online]. Available: https://github.com/google/skywater-pdk
\bibitem{ajayi2019} T. Ajayi et al., ``Toward an open-source digital flow: First learnings from the OpenROAD project,'' in Proc. of the 56th Annual Design Automation Conference (DAC), pp. 1-4, 2019.

\end{thebibliography}
\end{document}